\documentclass[noshowpacs,amsmath,twocolumn,superscriptaddress,8pt,aps,prb]{revtex4-1} % this article uses the Revtex 4-1 format
\usepackage{setspace}
\usepackage{amsmath}
\usepackage{breqn}
\usepackage{graphicx}
\usepackage[nearskip,margin = 0pt]{subfig}
\usepackage{verbatim} % for commenting
\usepackage{amsfonts}
\usepackage{amssymb}
\usepackage{epstopdf}
\usepackage{xcolor}
\usepackage[resetlabels,labeled]{multibib}
\DeclareGraphicsExtensions{.pdf,.eps,.png,.jpg,.mps}

\begin{document}
\title{Inverse-designed photonic circuits for fully passive, bias-free Kerr-based nonreciprocal transmission and routing}
\author{Ki Youl Yang$^{1,*}$, Jinhie Skarda$^{1,*}$, Michele Cotrufo$^{2,3*}$, Avik Dutt$^{1}$, Geun Ho Ahn$^{1}$, Dries Vercruysse$^{1}$, \\Shanhui Fan$^{1}$, Andrea Al\`{u}$^{2,3,4,5}$, and Jelena Vu\u{c}kovi\'{c}$^{1,\dagger}$\\
\vspace{+0.05 in}
$^1$E. L. Ginzton Laboratory, Stanford University, Stanford, CA 94305, USA.\\
$^2$Department of Electrical and Computer Engineering, The University of Texas at Austin, Austin, TX 78712, USA\\
$^3$Photonics Initiative, Advanced Science Research Center, City University of New York, New York, NY 10031, USA\\
$^4$Physics Program, Graduate Center, City University of New York, New York, NY 10016, USA\\
$^5$Department of Electrical Engineering, City University of New York, New York, NY 10031, USA\\
*These authors contributed equally to this work.\\
$^{\dagger}$Corresponding author: jela@stanford.edu}

 \maketitle

{\bf Nonreciprocal devices such as isolators and circulators are key enabling technologies for communication systems, both at microwave and optical frequencies. While nonreciprocal devices based on magnetic effects are available for free-space and fibre-optic communication systems, their on-chip integration has been challenging, primarily due to the concomitant high insertion loss, weak magneto-optical effects, and material incompatibility. We show that Kerr nonlinear resonators can be used to achieve all-passive, low-loss, bias-free, broadband nonreciprocal transmission and routing for applications in photonic systems such as chip-scale LIDAR. A multi-port nonlinear Fano resonator is used as an on-chip, all-optical router for frequency comb based distance measurement. Since time-reversal symmetry imposes stringent limitations on the operating power range and transmission of a single nonlinear resonator, we implement a cascaded Fano-Lorentzian resonator system that overcomes these limitations and significantly improves the insertion loss, bandwidth and non-reciprocal power range of current state-of-the-art devices. This work provides a platform-independent design for nonreciprocal transmission and routing that are ideally suited for photonic integration.}

\begin{figure*}[t!]
\centering
\includegraphics[width=\linewidth]{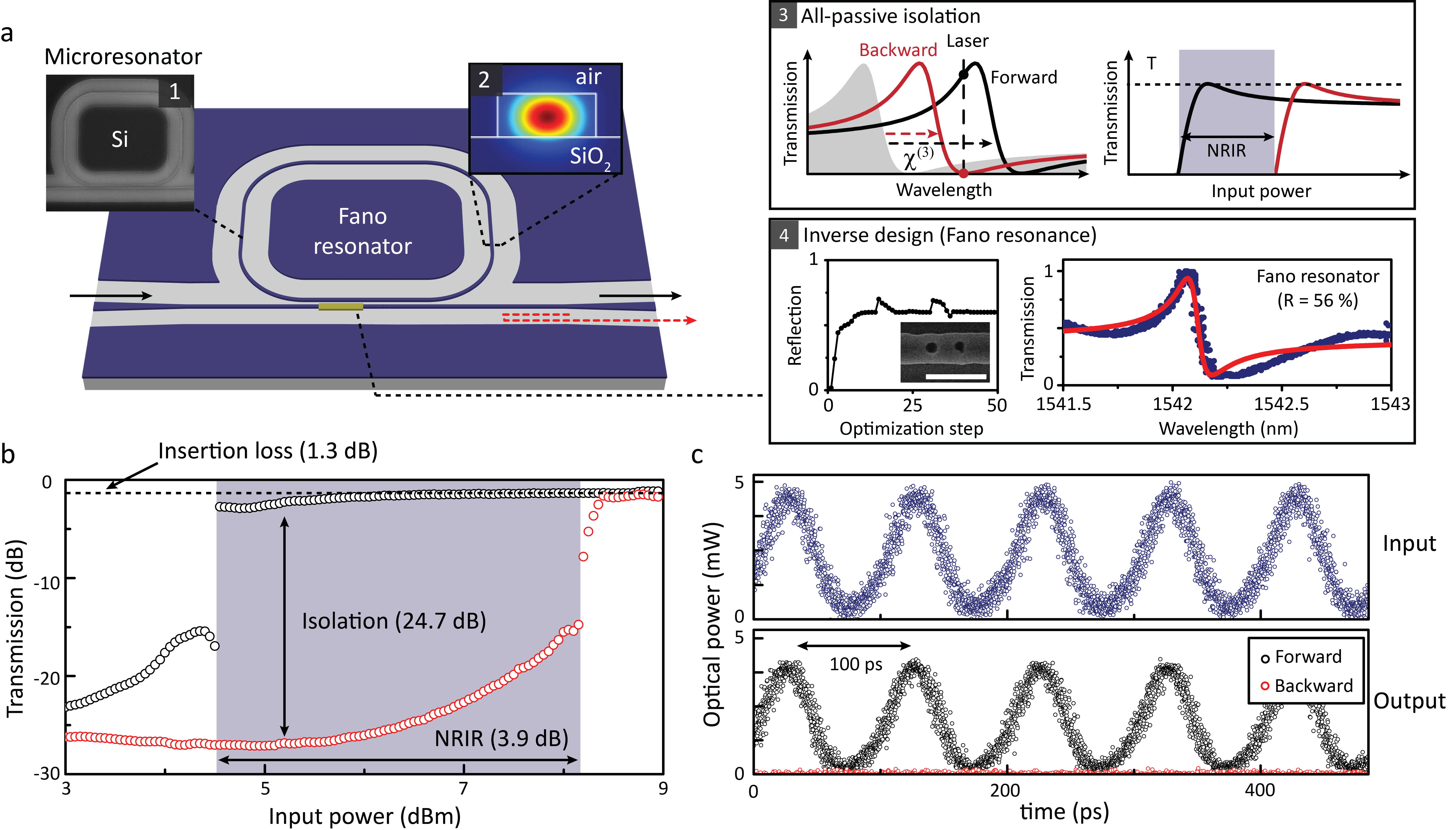}
\captionsetup{format=plain,justification=RaggedRight}
\caption{{\bf{All-passive nonreciprocal transmission using a silicon photonic resonator.}} (\textbf{a}) Schematic of the Fano nonreciprocal device. The silicon resonators (inset 1: Scanning electron micrograph (SEM) image of the resonator) only support a fundamental transverse-electric (TE) mode (inset 2: simulated spatial mode profile). A silicon waveguide is side-coupled to the race-track resonator, and the yellow box on the waveguide shows the location of the inverse-designed reflector used to synthesize Fano resonances. This inverse design area is offset with respect to the center of the cavity-waveguide coupling region in order to break the symmetry between the forward and backward direction coupling coefficients. As a result, forward and backward direction excitation with the same input intensity produce different shifts of the resonator resonant frequency, and thus of the transmission lineshape (inset 3). Inset 4: (Left) Inverse design optimization trajectory showing simulated device reflection versus optimization iteration, and SEM image of the final fabricated inverse-designed reflector (scale bar: 1 $\mu m$). (Right) Measured transmission spectrum of the fundamental TE mode in the silicon racetrack resonator with inverse-design-reflector. (\textbf{b}) Measured transmission versus input power to the Fano resonator in the forward (black) and backward (red) directions. The CW input is coupled to the input and output ports separately, and the transmitted signal is measured at the opposite port. (\textbf{c}) Transmitted signal trace of a 10 GHz modulated CW signal for excitation in either the forward or backward propagation directions (upper panel: input signal trace). Measured output power is calibrated with the chip-to-fibre interface efficiency.}
\label{fig:Fig1}
\end{figure*}

Nonreciprocal devices have become key components for enabling transmission and reception on the same communication channel with the rise of the new generation of cellular networks (5G) based on full-duplex radio-frequency communications. Similarly, as optical-frequency systems advance, it is becoming increasingly important to implement nonreciprocal devices in silicon photonic circuits. The ability to miniaturize nonreciprocal systems toward portable mobile scale and boost performance enables a wide range of applications in optical communications, signal processing, spectroscopy, and sensing. To date, integrated nonreciprocal devices have been demonstrated by using spatiotemporal modulation\cite{Yu:2009:NP,Lira:2012:PRL,Sounas:2018:NP}, magnetic bias\cite{Bi:2011:NP,Zhang:2019:arxiv,Huang:2016:JSQE}, Brillouin scattering\cite{Sohn:2018:NP,Kittlaus:2018:NP}, and optical nonlinearities\cite{Gallo:2001:APL,Fan:2012:Science,Mahmoud:2015:NC,Yu:2015:LPR,Peng:2014:NP,Chang:2014:NP}. Additionally, there have been important proposals and demonstrations in micro- and acoustic wave systems\cite{Fleury:2014:Science,Estep:2014:NatPhys}. While it is well established that linear non-reciprocity requires an external bias that breaks time-reversal symmetry, Kerr nonlinear devices eliminate this need and can be implemented in a monolithically integrated platform. Such devices can thus greatly simplify the fabrication steps as well as device design and operation. Despite dynamic reciprocity constraints\cite{Shi:2015:NP} on passive nonlinear devices under simultaneous excitation from both ports, silicon-based Kerr nonreciprocal devices are still attractive if outstanding challenges of integrated isolation and circulation devices\cite{Komljenovic:2018:IEEE} can be overcome. In particular, optical isolation in a large-scale integrated photonics has remained elusive mainly due to high insertion loss\cite{Fan:2012:Science,Fan:2013:OL,Huang:2016:JSQE,Lira:2012:PRL,Zhang:2019:arxiv}, narrow bandwidth\cite{Kim:2017:Scientific}, and scalability\cite{Kittlaus:2018:NP,Zhang:2019:arxiv}. The development of such a silicon-based chip-scale nonreciprocal device may lead to novel nonlinear devices for applications in optical communications and LIDAR\cite{Trocha:2018:Science}.

Here, we demonstrate all-passive, bias-free nonreciprocal transmission and routing of mode-locked pulse streams in silicon photonic systems. First, an asymmetric, single mode Fano resonance is implemented in a conventional microresonator side-coupled to a bus waveguide containing compact inverse-designed reflectors. Kerr nonlinearity enables self-biased and high-speed nonreciprocal transmission in the silicon microresonators, and the silicon photonic device exhibits extremely low insertion loss. Addition of a drop waveguide to the silicon Fano resonator enables nonreciprocal routing of a pulsed signal for a frequency comb based optical distance measurement. Importantly, it has been recently shown that any nonlinear single-resonator isolator is affected by a fundamental trade-off\cite{Sounas:2018:PRB} between the maximum forward transmission (T) and the nonreciprocal intensity range (NRIR) over which isolation can occur (Eq. \ref{eq:1}), such that a larger T can be obtained only at the expenses of a smaller NRIR, and vice-versa\cite{Sounas:2018:PRB}. By characterizing several single-resonator devices we provide, to the best of the our knowledge, the first experimental and systematic verification of this bound. For practical photonic applications, it is extremely important to relax this bandwidth-transmission trade-off.\cite{Komljenovic:2018:IEEE} As recently shown,\cite{Sounas:2018:NE} the single-resonator bound can be overcome by cascading two nonlinear resonators.
Following this approach, we exploit the large versatility of our inverse-design and fabrication process to implement a cascaded Fano-Lorentzian resonators device. This allows us to obtain near-unity transmission (T $>$ 99\%) over a NRIR $>6$ dB, largely beating the single-resonator bound. Beyond the direct demonstration and application of these nonlinear nonreciprocal devices as isolators and routers, this work illustrates a platform-independent design method to unlock novel functionalities in integrated nonlinear optics. Finally, we use the demonstrated nonreciprocal circuit to operate an on-chip LIDAR system, demonstrating the practical relevance of our result, which enables distance measurements at up to 60 m with a bias-free, fully passive non-reciprocal device. \\

%the device investigated in Fig. \ref{fig:Fig1} and also for other single-resonator devices as shown later in Fig. \ref{fig:Fig4}b. Importantly, our device design (see supplementary information section I) enables the single Fano isolators (Fig. \ref{fig:Fig1} and Fig. \ref{fig:Fig4}b) to operate close to the bound and thus obtain insertion loss lower than the records of the current state-of-the-art

\noindent\textbf{Device implementation.} The passive nonreciprocal Fano device consists of a silicon race-track resonator coupled to a silicon bus waveguide containing an inverse-designed reflector, as illustrated in Fig. 1a. The device layer has a thickness of 220 nm on 2-$\mu$m-thick silicon oxide layer, and the resonator was designed to operate on single mode at the wavelength of 1550 nm (inset 2). Using fabrication constrained inverse design\cite{Piggott:2015:NP,Piggott:2017:Scientific}, a partially transmitting element (PTE) was designed in the waveguide at the resonator coupling region to create a Fano lineshape in the cavity response function\cite{Fano:1961:PR,Fan:2003:JOSA,Fischer:2017:PRA,Yu:2014:APL}. The PTE is positioned with a slight offset (more details in the method section) from the center of the coupling region toward the input port, so as to create asymmetric coupling to the resonator from opposite ports\cite{Yu:2015:LPR}. Inset 4 (Fig 1a) shows the inverse design optimization trajectory and an SEM image of the fabricated PTE structure. Spectral measurements of a single microresonator with the PTE were performed by monitoring the waveguide transmission as the laser wavelength was scanned, and the right panel of the inset 4 presents a spectral scan near 1542 nm. Linewidth extracted from the fitted Fano curve\cite{Fano:1961:PR} shown in red gives a loaded Q factor of 7.3 $\times$ 10$^3$. \\

\noindent\textbf{Device characterization.}  
To validate the operation of the Fano nonreciprocal device, we measure the forward and backward transmission over a 10 dB range of continuous-wave (CW) input powers and evaluate the nonreciprocal intensity range (NRIR) (Fig. \ref{fig:Fig1}b). Following previous works,\cite{Sounas:2018:PRB,Sounas:2018:NE} we define the NRIR as the ratio of input powers from opposite propagation directions that leads to the threshold-like transmission transition. The forward and backward transmissions were measured without counter-propagating waves, and more experimental details are described in the method section. The maximum transmission contrast between the two directions is 24.7 dB (minimum: 13.4 dB), and the average contrast is 21.9 dB within the operating power range indicated by the shaded region. The insertion loss is 1.3 dB and the nonreciprocal operation range is 3.9 dB (operation power from 4.55 to 8.15 dBm loaded on the silicon waveguide). As mentioned above, the maximum forward transmission (T) and the NRIR are fundamentally bound\cite{Sounas:2018:PRB,Sounas:2018:NE} in a single Fano-resonance system by 
\begin{equation}
\label{eq:1}
T \leq  \frac{4 \cdot \textrm{NRIR}}{(\textrm{NRIR}+1)^2}.   
\end{equation} 
As a result, quasi-unitary transmission (i.e. very-low insertion loss) can only be achieved across a small range of impinging powers. We have experimentally verified that this bound holds for the device investigated in Fig. \ref{fig:Fig1} and also for many other single-resonator devices as shown later in Fig. \ref{fig:Fig4}b. Importantly, our device design enables the single Fano isolators to operate close to the bound (see Fig. \ref{fig:Fig4}b) and thus to obtain insertion loss lower than the records of current state-of-the-art devices\cite{Lira:2012:PRL,Huang:2016:JSQE,Fan:2012:Science,Fan:2013:OL,Yu:2015:LPR}. 

To demonstrate the applicability of this Fano nonreciprocal device to high-speed signal processing, we also characterize the forward and backward transmission with a 10 GHz modulated CW signal (Fig.\ref{fig:Fig1}c) generated using a Mach-Zehnder (MZ) modulator\cite{Yu:2015:LPR}. We use a peak power of approximately 5 mW ($\approx $ 7 dBm), corresponding to average energy for a period of about 250 fJ.
% I defined 
%Energy = P_max/ T *Integral_0^T [ cos(pi/T t)^2 dt] =
%=P_max/ T/2
%where T = 10ps
%and the operation power range per period (optical power) is approximately in the range of 100 - 1000 fJ. 
%This efficiency can be further improved by optimizing the cavity Q factor and mode volume. 
The forward transmission shows negligible waveform distortion, while the backward transmission is strongly suppressed (monitored by a digital communications analyzer). The high-speed operation is enabled by the fast relaxation speed of the resonator resonance (20 - 200 GHz) and Kerr nonlinearity. We note that our device operates in the power regime where free carrier effect is negligible\cite{Salem:2008:NP}. The threshold-like input-output power transfer function and high-speed operation capability, demonstrated in Fig.\ref{fig:Fig1}b and c, satisfy requirements for an all-optical regenerator\cite{Slavik:2010:NP,Li:2017:NP,Salem:2008:NP} and point to its potential usefulness not only for nonreciprocal devices but also for nonlinear signal processing in optical communications.   \\

\begin{figure*}[t!]
\centering
\includegraphics[width=\linewidth]{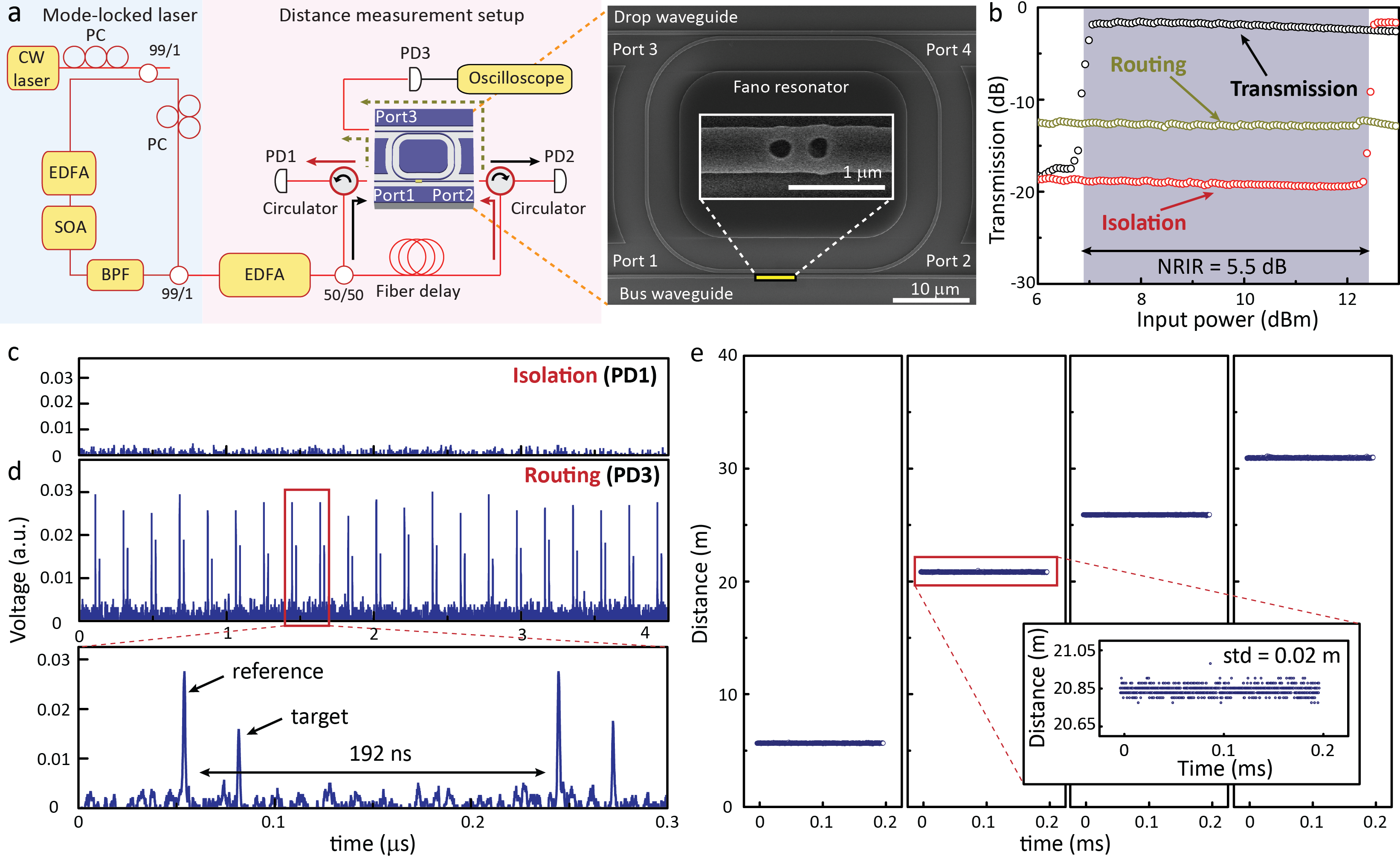}
\captionsetup{singlelinecheck=no, justification = RaggedRight}
\caption{{\bf{Nonreciprocal pulse routing and optical distance measurement}} (\textbf{a}) Schematic of the pulsed laser generation setup (left) and optical distance measurement setup (right). A CW laser is polarization-controlled (PC) and split using 99/1 coupler to pump the fibre loop resonator. An erbium-doped fibre amplifier (EDFA) and a semiconductor optical amplifier (SOA) are used as the gain media of the fibre laser\cite{Tamura:1993:OL}, and a bandpass filter (BPF) selects the central frequency of the output pulse. For the distance measurement, the pulse stream is split into two paths: the reference path is directly connected to device port 1, while the target path includes a fibre delay prior to port 2. The pulse streams of both paths are routed to port 3 and detected at photodetector 3 (PD3) for a time-of-flight measurement. PD2 monitors the forward transmission through the device. PD1 records the transmission of the target path pulse stream from port 2 to port 1, thus monitoring the device's ability to isolate the pulsed laser input source (Inset: SEM image of device). (\textbf{b}) Transmission versus input power for a CW input through three different propagation paths corresponding to transmission ($S_{12}$), isolation ($S_{21}$), and routing ($S_{23}$), where $S_{ij}$ indicates transmission from port i to port j. (\textbf{c}) The electrical intensity trace of PD1 during optical ranging, demonstrating isolation of the pulsed laser input source. (\textbf{d}) Representative electrical intensity trace from PD3 showing reference and target pulses containing ranging information. The lower panel shows the zoom-in of the electrical intensity trace over two pulse periods. The pulse period is approximately 192 ns. (\textbf{e}) The measured distance between the reference peak and target peak versus time (data acquisition time: 200 $\mu$s) for fibre delay path length of 5, 20, 25, and 30 m respectively from left to right. Inset: Zoom-in of measured distance trace showing the standard deviation of 0.02 m, demonstrating the stability of this optical distance measurement. The range ambiguity in this measurement is approximately 40 m in silica fibre.}
\label{fig:Fig2}
\end{figure*}

\begin{figure*}[t!]
\centering
\includegraphics[width=1.0\linewidth]{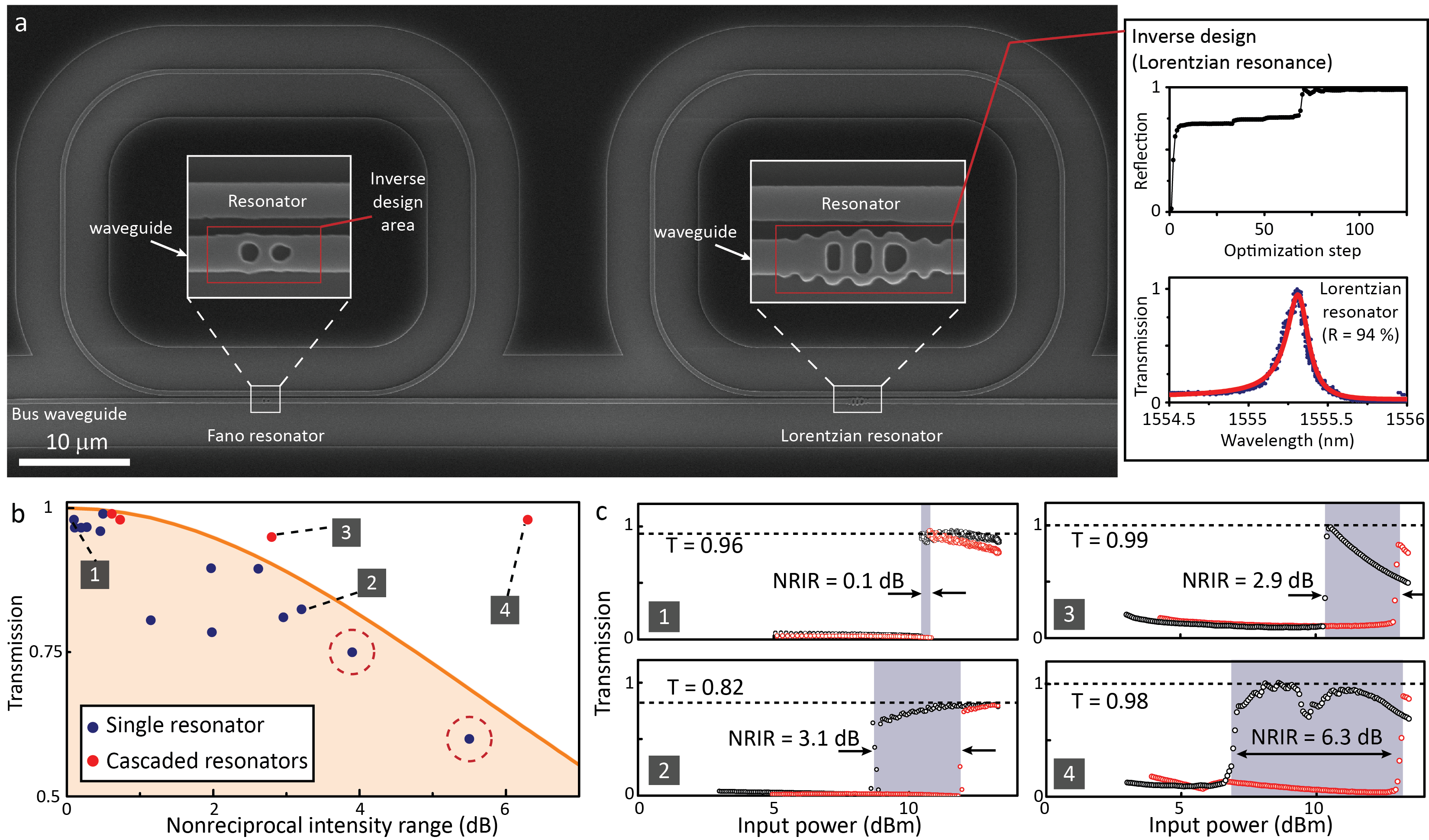}
\captionsetup{format=plain,justification=RaggedRight}
\caption{{\bf{Broadband isolation operation using cascaded nonlinear resonators}} 
(\textbf{a}) SEM image of cascaded Fano/ Lorentzian resonators implemented on silicon-on-insulator platform. The zoom-in images show inverse-design-reflectors on the silicon waveguide at the resonator-waveguide coupling region. The inset shows the inverse-design of the 94 \% reflector used to implement a Lorentzian resonator (optimization trajectory, SEM of the final design, and obtained resonance lineshape). The Fano resonator is implemented using the same inverse-designed reflector shown in the inset of Fig. 1a. (\textbf{b}) Forward transmission versus nonreciprocal intensity range for single and cascaded resonators. The shaded region corresponds to the theoretical bound on single resonator operation\cite{Sounas:2018:PRB,Sounas:2018:NE}, and the dots correspond to measurements of single resonator (blue) and cascaded resonator (red) systems. The dots with red-dashed circles are measurement results of the devices presented in Fig. 1 and 2. (\textbf{c}) Measured transmission versus input power of single resonators (Panel 1, 2) and cascaded resonators (Panel 3, 4) in the forward (red) and backward (black) directions.}
\label{fig:Fig4}
\end{figure*}

\noindent\textbf{Demonstration of optical ranging measurement based on nonreciprocal transmission and all-optical routing.} While Kerr nonlinear isolators are fundamentally limited by time-reversal symmetry and thermodynamic considerations, their passive, magnet-free, bias-free, simple architecture makes them particularly appealing for integrated photonics. We show how these devices can be used in a range of important photonic systems operating with pulsed signals, where the forward and backward ports are not simultaneously excited (see Fig. \ref{fig:FigS1} for additional details). Despite recent advancements in photonic integration such as large-scale phased array\cite{Sun:2013:Nature} and microcomb sources\cite{Trocha:2018:Science} that are a promising step towards miniaturization of LIDAR systems, other chip-based components like isolators and circulators are still required in a full system to harness the potential of these approaches. As a demonstration of how our silicon device is useful for integrated LIDAR systems, we perform precise and reliable optical distance measurements using a frequency comb as the source and our device as both the on-chip optical isolator and router. A pulse routing waveguide is added to the Fano nonreciprocal device (device schematic in Fig. \ref{fig:Fig2}a) to enable the device to guide the pulses from both the pump laser and the detection target to the same photodetector while protecting the pump laser from the reflected pulse.

The experimental schematic used for the optical ranging measurement is illustrated in Fig. \ref{fig:Fig2}a. For mode-locked pulse generation, we use a fibre ring resonator consisting of an erbium-doped fibre amplifier (EDFA), a semiconductor optical amplifier (SOA), polarization controller (PC), and bandpass filter (BPF). The 5 MHz pulse repetition rate allows measuring distances up to 60 m, and the central frequency of the pulse can be adjusted to the Fano isolator operation range using the BPF. The generated pulse streams are split by a 50:50 fibre-based coupler, and one part is directly sent to device port 1. The other part of the pulse stream is first passed through a fibre delay line (physical path length $\sim$ 5, 20, 25, 30 m) and then sent to port 2. The pulse stream coupled to port 2 does not propagate to port 1 because of the Fano isolator (as checked by monitoring PD1, Fig. \ref{fig:Fig2}c), while the pulse stream from port 1 is transmitted to port 2 of the device (monitored at PD2). The device routes the pulse stream from port 2 to port 3 in the other waveguide coupled to the Fano resonator (Fig.\ref{fig:Fig2}b), where it combines with the input pulse stream from port 1 to generate an electric signal trace of dual pulse streams (Fig. \ref{fig:Fig2}d) at PD3. A zoom-in view of the trace (lower panel of Fig. \ref{fig:Fig2}d) shows the reference and target peaks within two periods of 192 ns, and the time interval between reference and target peaks is calculated for each period and converted to the distance scale. Fig. \ref{fig:Fig2}e plots distance versus time where the time increment is the reference pulse period, showing the stability of the distance measurement. To extend this distance measurement setup into a LIDAR system, the device can be equipped with an integrated microlens\cite{Trocha:2018:Science,Dietrich:2018:NP}, metalens\cite{Arbabi:2015:NC}, and phased array\cite{Sun:2013:Nature} at port 2 to emit the pulse stream from port 1 towards the target and to receive the reflected pulse streams. Received pulse in a typical LIDAR system is a low-intensity signal hence back-reflection at the isolated port and pulse emitter of the signal is also weak, making it practically suitable for a precise distance measurement. Our Fano isolator-router device can protect the pulsed laser source from the reflected pulse streams for stable system operation while routing light from these reflected pulses to the other waveguide. In this measurement, the reference and reflected pulses do not arrive on our device at the same time and hence bypass the dynamic non-reciprocity constraint\cite{Shi:2015:NP}. Such operation, in principle, can be also realized using active time-gated switching but high-speed operation is challenging\cite{Cheng:2018:OE}. Furthermore, we expect that reflected pulses can be isolated even under simultaneous excitation if another isolator is connected in series with a time delay corresponding to half the pulse period. Based on these findings, our experiment demonstrates the viability of the realized passive isolator and router to act as an essential component in a fully integrated chip-scale LIDAR system.\\

\noindent{\bf Broadband operation based on cascaded nonlinear resonances.}
Unlike linear active isolators that can potentially work under any input power, nonlinear passive devices based on a single resonator can lead to isolation only over a limited range of signal powers (NRIR). In fact, as already mentioned above, single-resonator devices are affected by a fundamental trade-off between the maximum achievable forward transmission and the NRIR \cite{Sounas:2018:PRB}. Here, we provide the first experimental verification of this fundamental constraint by characterizing several single-resonator isolators. The different designs are obtained by varying the reflectivity of the inverse-designed coupling element and the cavity-waveguide gaps in the device in Fig. \ref{fig:Fig1}a. The blue dots in Fig. \ref{fig:Fig4}b show the measured forward transmissions and NRIRs for each single-resonator device, and the shaded region corresponds to the theoretical bound in Eq. \ref{eq:1}. All the investigated single-resonator devices are clearly constrained by the bound. As an example, Panel 1 and 2 of Fig. \ref{fig:Fig4}c show transmission versus input power for single resonators in the forward and backward directions. These results clearly illustrate that a highly transmissive (T = 0.96) device is affected by a narrow operating power range ($\sim$ 0.1 dB), while a less transmissive (T = 0.82) device can operate on a broader power range (3.1 dB). If unitary transmission is required, the operation range for a single resonator device shrinks to zero. 

Increasing the power bandwidth of these isolators while keeping quasi-unitary transmittance is fundamental for their use in many photonic applications \cite{Komljenovic:2018:IEEE}. It was shown in \mbox{Ref. [\!\!\citenum{Sounas:2018:NE}]} that two cascaded nonlinear resonators, interleaved by properly chosen delay lines, can overcome this constraint and realize broadband isolation beyond the bound in Eq. \ref{eq:1}. We implemented the system proposed in \mbox{Ref [\!\!\citenum{Sounas:2018:NE}]} consisting of cascaded Fano-Lorentzian resonators (Fig.\ref{fig:Fig4}a). The Lorentzian response is obtained using the same race-track resonator configuration, and optimizing a highly reflective waveguide structure in the coupling region in such a way that Lorentzian instead of Fano response is achieved (Inset of Fig.\ref{fig:Fig4}a shows the inverse-design of this structure). The Fano resonator is nominally identical to the device characterized in Panel 1 of Fig.\ref{fig:Fig4}c. A key element to overcome the single-resonator bound is the ability to control the phase delay between the Fano and the Lorentzian resonator.\cite{Sounas:2018:NE} In our device the phase delay is set lithographically by the physical length of the input waveguide that couples to both resonators (red arrow in Fig.\ref{fig:Fig4}a).  We characterize several Fano-Lorentzian cascaded systems, with various phase delays, and compare their isolation performance with the single Fano devices. The red dots in Fig. \ref{fig:Fig4}b show the transmission versus NRIR of cascaded resonators measured at the same wavelength as the single resonator systems. For selected phase delays, we obtain a clear breaking of the single-resonator bound. Panel 3 and 4 of Fig. \ref{fig:Fig4}c show the transmission versus input power for two cascaded resonator systems that break the single-resonator. The measured data show high forward transmission of 95 - 99 \% (0.04 - 0.22 dB insertion loss) with broader operating power ranges (6.3 dB) than allowed by the single resonator limit. We note that in this all-passive device experiment no attempt to actively control resonance wavelengths and phase delay was made, but in the future thermo- or electro-optic control can be introduced to further optimize the performance. These experimental results demonstrate a practical and feasible solution to overcome the  power range constraints of nonlinear devices. In addition, the capability to  precisely tailor the coupling between nonlinear resonators and a common reservoir could allow, for example, the realization of arrays of nonlinear elements with self-induced topological protection in the optical domain\cite{Hadad:2018:NE}.\\

\noindent{\bf Conclusion}\\
In summary we have demonstrated that Kerr nonlinear resonators can be used to achieve fully passive, bias-free nonreciprocal transmission and routing in standard silicon photonic systems, of special interest for chip-scale LIDAR. The system architecture was optimized by photonics inverse design. In order to increase the non-reciprocal intensity range while preserving high transmission, a cascaded system of a Fano and Lorentzian resonator was studied. Unitary forward transmission as well as broad operating power range were demonstrated to illustrate the achievement of essential, but previously elusive, functionalities required for application of these nonreciprocal devices to practical photonic systems. Although such nonlinearity-based devices are mainly useful for pulsed or periodic source applications due to dynamic reciprocity constraints \cite{Shi:2015:NP}, the non-reciprocal signal transmission and routing demonstrated in this work open the door to many other practical applications of immediate interest in photonics. These nonlinear photonic systems provide novel functionalities for advanced optical communications, nonlinear signal processing, spectroscopy, and all-optical circuits in a silicon-nanoelectronics-compatible photonic platform\cite{Atabaki:Nature:2018}. \\

\noindent{\bf Methods}\\
\noindent\textbf{Inverse design} Stanford Photonics Inverse Design Software (SPINS)\cite{SPINs}, based on the previously described fabrication-constrained inverse-design methodology\cite{Piggott:2015:NP,Piggott:2017:Scientific}, was used to inverse-design waveguide reflectors for target reflection = 60, 80, 99 $\%$ for 1550 nm fundamental-mode TE-polarized light. These reflectors were designed inside the 500 nm width input-waveguide with a design area of 500 nm $\times$ 2000 nm and a minimum feature size constraint of 120 nm.\\

\noindent\textbf{Characterization} Transmission measurements of the nonlinear isolators were performed by coupling a continuous-wave (CW) laser through a single-mode fibre onto the chip via grating couplers\cite{Sapra:2019:IEEE} and monitoring the transmitted power through the waveguide-coupled device in the forward and backward directions while sweeping the laser input power (measurement wavelengths: 1541.4 nm in Fig.\ref{fig:Fig1}, 1533.87 in Fig.\ref{fig:Fig2}, 1533.87, 1541.4, 1547.5, 1552.1, 1556.6, 1562 nm in Fig.\ref{fig:Fig4}b, 1562 nm in panel 1, 3, and 4 of Fig.\ref{fig:Fig4}, 1547.5 nm in panel 2 of Fig.\ref{fig:Fig4}). The transmission values are normalized by performing a transmission measurement of a single waveguide with grating couplers, and the propagation direction was changed by implementing a fibre switch. It is important to note that the transmissions in both directions were measured without counter-propagating waves in the device, and the counter case (dynamic reciprocity\cite{Shi:2015:NP}) is discussed in other work\cite{Cotrufo:2019}. \\

\noindent\textbf{Resonator design parameters} Waveguide-cavity gap = 200 nm, Cavity-waveguide interaction length = 18.3 $\mu$m, PTE offset relative to midplane of a single Fano resonator = 2.28 $\mu$m, delay lengths between Fano and Lorentzian resonators = 44 and 43.75 $\mu$m in panel 3 and 4 of Fig.\ref{fig:Fig4}c, respectively. \\

\noindent\textbf{Data availability} The data that support the plots within this paper and other findings of this study are available from the corresponding author upon reasonable request.

\medskip

\noindent\textbf{Acknowledgment}
\noindent We acknowledge insightful discussion with D. Sounas, W. Bogaerts, and D.A.B. Miller, and also thank technical advise from C. Langrock, B. Buscaino, N.V. Sapra and J.M. Kahn. The silicon devices were fabricated in the Stanford Nanofabrication Facility and the Stanford Nano Shared Facilities. K.Y.Y. acknowledges Quantum and Nano science and engineering postdoctoral fellowship, J.S. acknowledges the National Science Foundation Graduate Research Fellowship (Grant No. DGE-1656518), and M.C. is supported by a Rubicon postdoctoral fellowship by The Netherlands Organization for Scientific Research (NWO). This work is funded by the Air Force Office of Scientific Research under the AFOSR MURI program (Award No. FA9550-17-1-0002) and the Gordon and Betty Moore Foundation (GBMF4744, GBMF4743). We thank Gernot Pomrenke and AFOSR MURI program management team for discussions throughout the project. \\

\noindent\textbf{Author contributions.}
K.Y.Y., J.S., M.C., A.A., and J.V. conceived the experiments. K.Y.Y., J.S., and M.C. designed the device. K.Y.Y., J.S. fabricated and tested the devices with assistance from M.C., A.D., G.H.A., and D.V. K.Y.Y., A.D., and J.S. conducted optical ranging measurement with assistance from M.C., and D.V. All authors analyzed the data and contributed to writing the manuscript. J.V. and A.A. supervised the project.\\

\noindent\textbf{Competing interests.}
The authors declare they have no competing financial interests.

\bibliography{Reference}
\clearpage

\onecolumngrid 

\appendix 
\renewcommand{\thefigure}{S\arabic{figure}}
\renewcommand{\thesection}{\Roman{section}}
\setcounter{figure}{0} 
\section* {Supplementary Information to\\Inverse-designed photonic circuits for Kerr nonreciprocal transmission and routing}
\vspace{-0.15 in}
\noindent Ki Youl Yang$^{1*}$, Jinhie Skarda$^{1*}$, Michele Cotrufo$^{2,3*}$, Avik Dutt$^{1}$, Geun Ho Ahn$^{1}$, Dries Vercruysse$^{1}$, \\Shanhui Fan$^{1}$, Andrea Al\`{u}$^{2,4,5,6}$, and Jelena Vu\u{c}kovi\'{c}$^{1,\dagger}$
\vspace{0.1 in}\\
\noindent
$^1$E. L. Ginzton Laboratory, Stanford University, Stanford, CA 94305, USA.\\
$^2$Department of Electrical and Computer Engineering, The University of Texas at Austin, Austin, TX 78712, USA\\
$^3$Photonics Initiative, Advanced Science Research Center, City University of New York, New York, NY 10031, USA\\
$^4$Physics Program, Graduate Center, City University of New York, New York, NY 10016, USA\\
$^5$Department of Electrical Engineering, City University of New York, New York, NY 10031, USA\\
$^*$KYY, JS, MC contributed equally to this work.\\
$\dagger$Correspondence and requests for materials should be addressed to jela@stanford.edu.\\

\section{Characterization of inverse-design-reflectors}

\begin{figure*}[h!]
\centering
\includegraphics[width=0.5\linewidth]{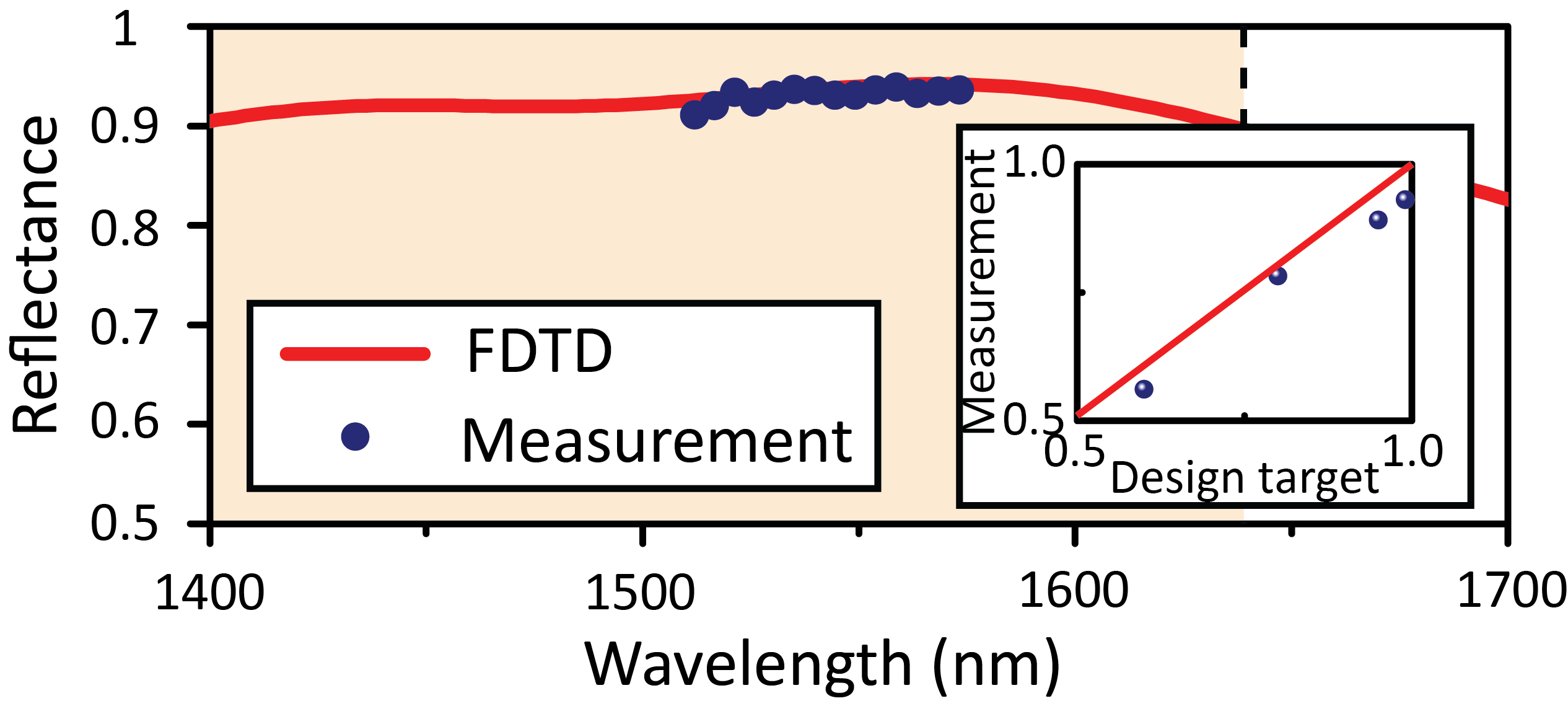}
\captionsetup{format=plain,justification=RaggedRight}
\caption{{\bf{Optical characterization of inverse-designed structures}} Measured reflectance spectrum of the inverse-designed structure in the inset of Fig.\ref{fig:Fig4}a. The measured spectrum (blue dots) agrees with numerical simulated spectrum of the fabricated structure (red). Inset: measured reflectances at 1550 nm of inverse designed structures versus design targets (60, 80, 95, 99 \%). Fano resonators in Fig.\ref{fig:Fig1}-\ref{fig:Fig4} (main text) used the inverse-designed structures of 60 and 80 \%.}
\label{fig:FigS3}
\end{figure*}

The reflectance of inverse-designed structure was characterized by measuring finesse of Fabry-Perot resonators consisting of single mode waveguide and identical reflectors at both ends\cite{Yariv}. Waveguide propagation loss is separately estimated using both cut-back method\cite{Vlasov} and Q factor of micro-ring resonators. Numerically simulated results (finite difference time domain; FDTD) shows a fair agreement with the measured spectrum. Inset shows measured reflectance of final design (1550 nm) as a function of initial optimization target.  

\section{Operation of a Fano device -- spectral bandwidth, operation power}
\begin{figure*}[h!]
\centering
\includegraphics[width=0.5\linewidth]{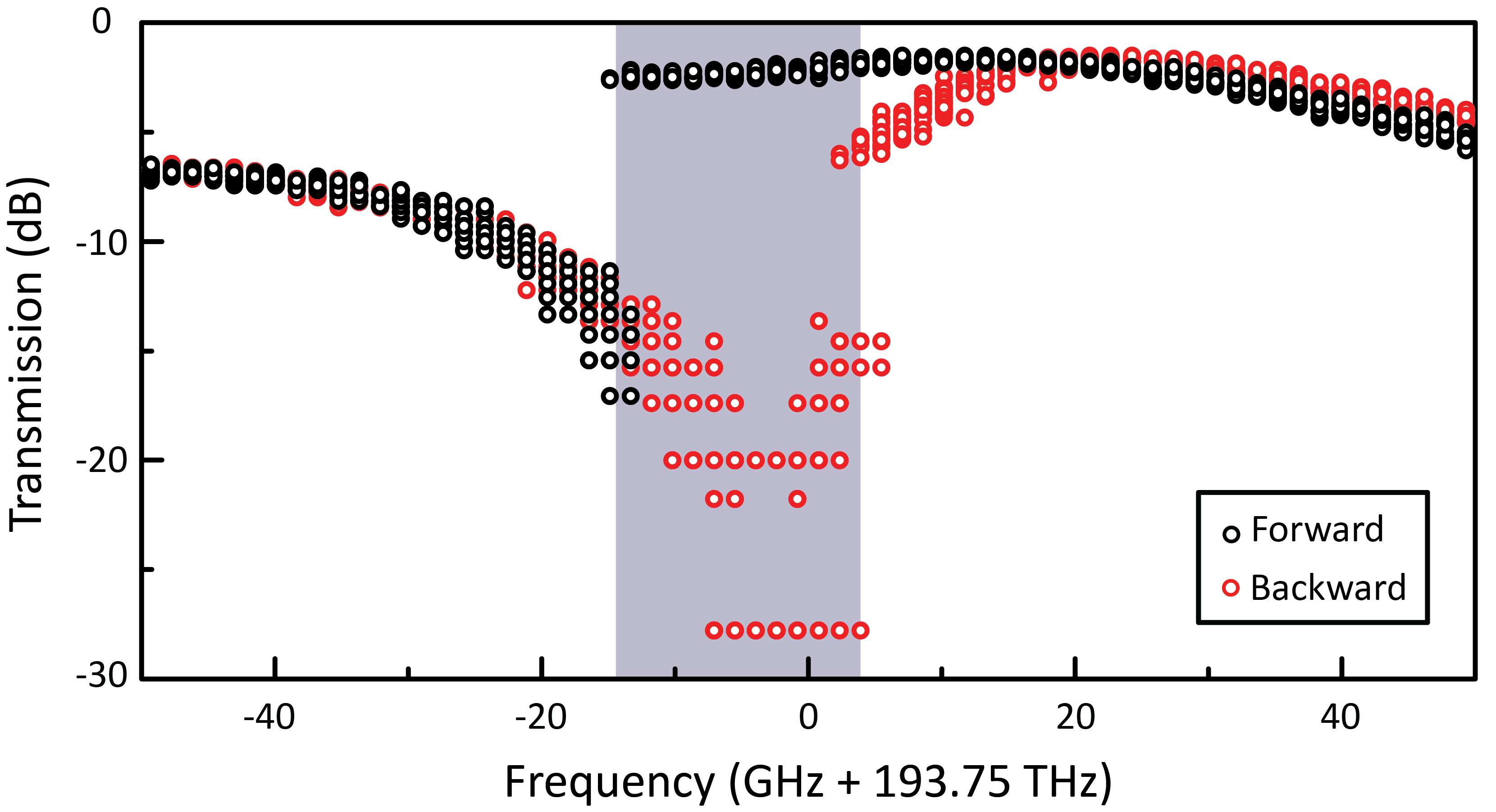}
\captionsetup{format=plain,justification=RaggedRight}
\caption{{\bf{Spectral bandwidth of a Fano device operation.}} Measured transmission in forward (black dots) and backward directions (red dots) versus laser wavelength.}
\label{fig:FigS2}
\end{figure*}

Fig.\ref{fig:FigS2} shows transmission of CW light in forward and backward directions as a function of wavelength (single Fano resonator with 80 \% reflectance and 2.28 $\mu$m offset relatived to midplan of resonator). Transmission was measured at different laser detuning with respect to the Fano resonance, and input CW power is constant throughout the entire measurement (14 dBm loaded in the silicon waveguide). The transmission in both directions were measured without counter-propagating waves in the device. 

\section{Nonreciprocal transmission of pulsed signal in an asymmetric Fano resonator}

\begin{figure*}[h!]
\centering
\includegraphics[width=\linewidth]{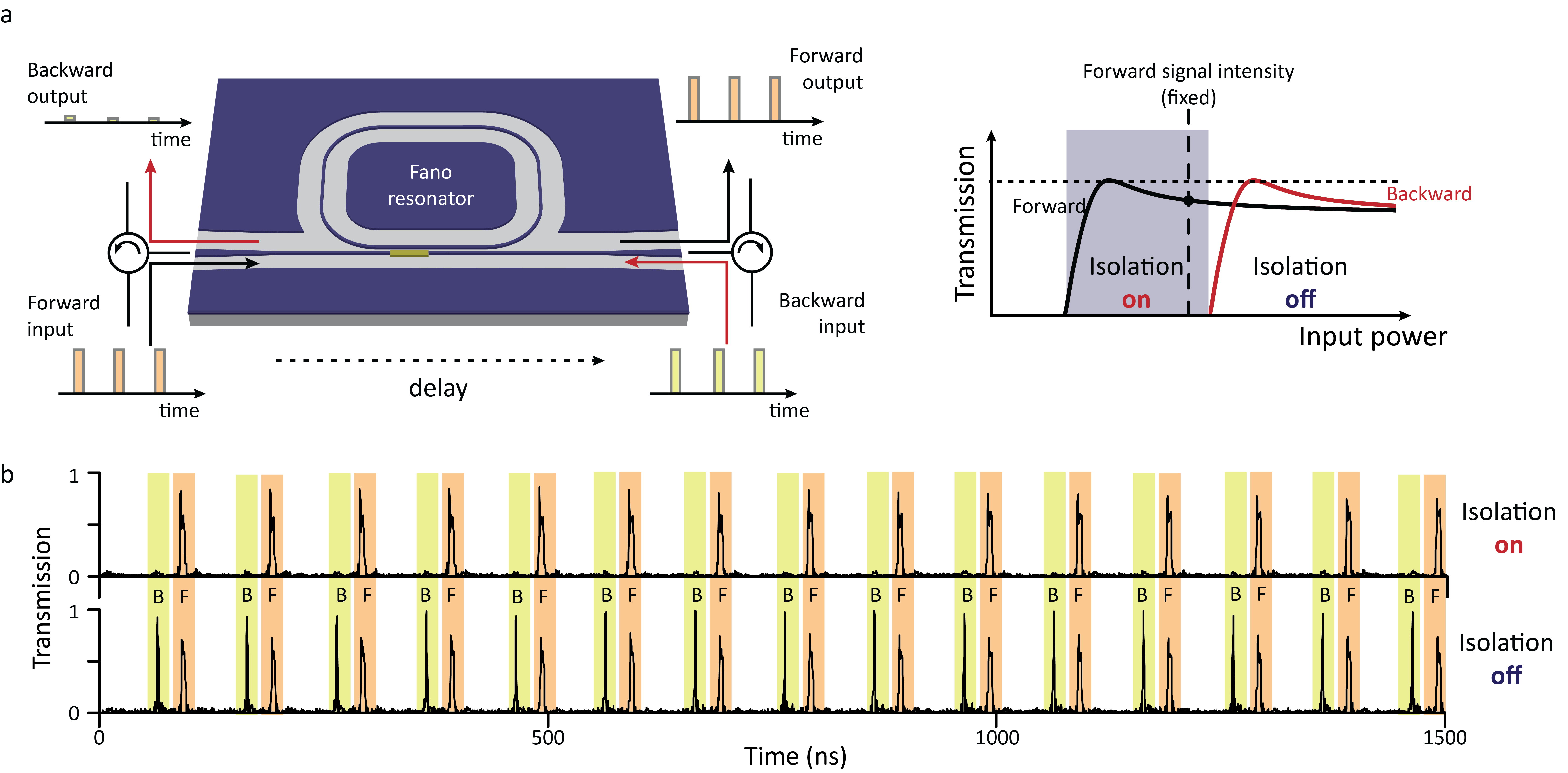}
\captionsetup{format=plain,justification=RaggedRight}
\caption{{\bf{All-passive isolation of pulsed signal.}} (\textbf{a}) Setup for nonlinear isolation experiment. The pulsed signal is simultaneously coupled to in- and output ports with time delay, and the transmitted signals were routed to a fast photodetector by optical circulators. (\textbf{b}) Transmission of pulsed signal in forward and backward directions at different backward input intensities.}
\label{fig:FigS1}
\end{figure*}

The experimental demonstration of nonlinear isolation is illustrated in Fig. \ref{fig:FigS1}a. Pulsed signal (repetition rate: 10 MHz) propagates to both forward and backward directions in the silicon waveguide with slightly different time delay, as a result overlap between forward and backward propagating pulses can be avoided. Transmission of pulsed signal in both directions is routed to photodetector and recorded in real time. To validate the operation of isolation, we initially launch pulse trains into waveguide in forward direction with signal intensities that make device highly transmissive to the same propagation direction, and read the transmission of backward propagating pulses in various backward input intensities. Upper panel of Fig. \ref{fig:FigS1}b shows transmission of pulse streams in forward and backward directions with lower backward input intensity while the lower panel shows the results with higher backward input intensity.

\end{document}